# An In-Field Programmable Adaptive CMOS LNA for Intelligent IOT Sensor Node Applications

Maryam Shafiee, Sule Ozev

**Abstract-** As the Internet of Things (IOT) is growing rapidly, there is an emerging need to facilitate development of IOT devices in the design cycle while optimized performance is obtained in the field of operation. This paper develops reconfiguration approaches that enable post-production adaptation of circuit performance to enable RF IC re-use across different IOT applications. An adaptable low noise amplifier is designed and fabricated in 130nm CMOS technology to investigate the post-production reconfiguration concept. A statistical model that relates circuit-level reconfiguration parameters to circuit performances is generated by characterizing a limited number of samples. This model is used to predict the performance parameters of the device in the field. The estimation error for LNA performance parameters are obtained in the simulation environment as well as chip measurements.

## I. INTRODUCTION

Internet of Things (IOT) is rapidly being integrated into our daily lives in diverse applications ranging from health care to home automation, energy management to environmental monitoring [1-4]. IOT devices have their own application-specific requirements which demands for multi-standard multi-mode transceivers. Conventionally the IOT interconnected objects are realized by existing commercial off-the-shelf components (COTS) which are designed and optimized for certain communication standards or a specific use [5-6]. However, using separate radios is power hungry, costly and the interconnections are not optimized with regards to an application-specific requirement. On the other hand, using customized radio transceiver ICs for each specific application is not practical due to high overall product costs (OPC). A cost-effective solution is a single radio transceiver adaptable to localized IOT applications. Reconfigurability enables power optimal interconnections within IOT devices. The reconfigurable transceiver is tweaked on the spot based on application-specific requirements such as gain, linearity, BER, etc. Having such adaptable RF ICs in the marketplace will enable IoT developers to optimize the overall system performance without having a deep RF design experience and without having to incur the cost of taping out an entirely new design for each product.

Wireless transceivers are increasingly complex and highly integrated systems which makes them more susceptible to process variations. During circuit design, a designer's primary goal is to meet circuit specifications under given process variations. In doing so, designers spend significant effort to minimize the effect of process variations or in other words, they try to de-sensitize their design with respect to process variations. While designers strive for process robustness at nominal operating conditions, such as supply voltage, noise, temperature, same robustness is generally difficult to maintain over a large variation in operating conditions. By modifying these operating conditions during testing, we can increase sensitivity to process parameters. There are some form of post-production calibration and reconfiguration for RF devices [7-9]. The calibration is realized by built-in tuning knobs allowing for trade-offs between RF specifications. In general, RF circuits are designed to include calibration hooks in bias or passive components to meet target specifications. Different calibration mechanisms are introduced in the literature in the form of bias current, bias voltage, and passive bank adjustments [10-13]. In general, Analog inputs are not desirable in RFIC design. Besides, these methods typically provide limited calibration space.

As opposed to post-production calibration which is optimizing the device with respect to a single application, in this paper, we develop existing mechanisms to reconfigure the RF device for optimized performance with respect to multiple IOT applications. We will exploit existing calibration approaches and mechanisms and enhance them to further sensitize the circuit to process/layout variations. This sensitization is expected to spread and shift the circuit performance distribution over process variation, as illustrated in Fig.1.

Our technique proposes a programmable device which adapts to different situations depending on the application requirements. Its performance is optimized in the field with regards to required specifications e.g. distance, power consumption, BER, data rate, etc. We propose to use statistical models to capture the correlations among measured performance parameters and reconfiguration modes. We employ machine learning technique to capture these non-linear correlations and predict the probability distribution of a target parameter based on measurements of correlated parameters. We will show that decision-making based on the prediction algorithm instead of iterative testing of all possible reconfiguration schemes, saves us time and facilitates fast in-field adaptation of the device. We have demonstrated the concept by designing an LNA with built-in tuning knobs. The tuning knobs are carefully designed to provide independent adjustment of important performance parameters such as gain and linearity. Minimum number of switches are used to provide the desired tuning range without a need for an external analog input.

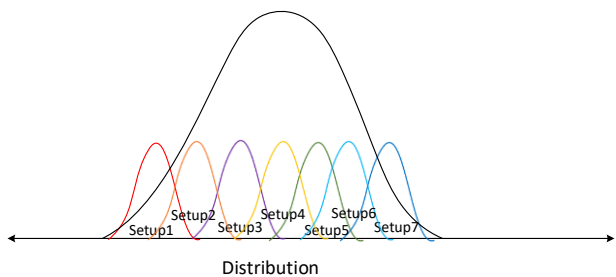

Fig.1. Sensitization over process variation using different calibration setup

In the following sections, the proposed performance prediction and optimization algorithm is discussed. The LNA architecture and selection of the tuning knobs are demonstrated. Simulation and chip measurement results are further discussed.

## II. PRIOR WORK

Adaptive wireless receivers have been explored in several published articles. Majority of these reconfigurability methods use power supply, bias current, bias voltage and passive bank adjustments [8,10,11,14,15]. These techniques employ fine and continuous tuning using analog control signals generated by simple low-speed digital to analog converters (DACs). However, DACs are power hungry and require notable dedicated silicon area. Besides, the DAC settling time and conversion rate limits the critical in-field adaptation pace.

Due to process variations, the performance parameters of the DUT are not fixed for all parts. This necessitates an in-field verification of reconfiguration state with respect to the target performance. The reconfiguration and verification procedure can be statistical-based or iterative. The iterative procedure makes a measurement with each adjustment of the tuning knobs until the target performance is achieved [8,11,15]. This procedure is time consuming. The statistical based method on the other hand, develops a nonlinear prediction model to adjust the tuning knobs. This method is fast since it is a one-time calibration procedure which requires a training set [10,16,17]. This methodology can be used for on-chip self-testing and calibration. Statistical modeling allows for easier model generation by relying on machine learning [18]. In [10], the nonlinear prediction model is developed using successive learning process based on MARS algorithm and Monte-Carlo samples. A self-generated sine test signal is measured for each DUT and is used as the input to the statistical model for prediction results. The DUT calibration knobs are adjusted using the performance curves of the DUT. The performance curve represents the relation between performance parameter and the knob value of a golden DUT. Since the knob values are continues, obtaining this curve for each DUT is not practical for in-field calibration. Hence, it only obtained for one sample (Golden DUT) and is used for tuning the rest of samples. Therefore, it does not take into account the process variation between DUTs for knob value selection.

None of the discussed existing approaches, address an automated fully digital reconfiguration scheme which sensitizes the tuning range to process variation.

## III. PROPOSED POST-PRODUCTION OPTIMIZATION FLOW

Our proposed reconfiguration approach uses transistor sizing and bias control. It uses coarse tuning of performance parameters which is realized by only switches and is fully performed in digital. Hence it is low cost and low overhead. As it is shown in Fig. (1), the performance distribution over process variation for each switch combination has overlap with others. This bring a level of uncertainty which necessitates the verification of the performance of the DUT at potential switching combinations. To decide on the optimum switch combination, instead of lengthy testing of the performance at each potential reconfiguration state, we have used a fast statistical-based prediction procedure using neural network algorithm. Our approach predicts the performance parameters of all switch combinations of the DUT and based on the prediction result, selects the switch combination closest to the target. Therefore, unlike previous works, the selection of the switch combination, takes into account the DUT to DUT variations.

Using simulations, we will construct local statistical models that relate circuit-level calibration parameters to circuit performances. We conduct Monte-Carlo simulations to obtain parameter profiles. These profiles will help in selecting and guiding the neural network training process. It is important to select the inputs to the neural network that are highly correlated to the target and these correlations are altered by potential process and circuit modifications. As it is depicted in the flow chart of Fig.2(a), a Monte-Carlo generated training set is input to the learning algorithm to form the statistical model of the system in all available states. Using neural network algorithm, the performance parameters are predicted for each switch combination while input parameters are assumed to be known at one or more combinations (e.g. at combination where all switches are off). Eventually, for a specified target performance and based on predicted performance parameters, the optimized operation is obtained by setting the correct digital code to the calibration network.

For better understanding of the neural network algorithm a diagram is shown in Fig.2(b). Set {p} represents the demanded specifications to characterize the device. Set {P} at switch combination zero, i.e. all switches off, is input to the neural network. The neural network learning process predicts the set {P} for the rest of the switching combinations. Due to overlap between performance distribution of switching combinations, it might be required to redo the prediction with two known inputs if it results in a closer to target combination selection. This means adding another measurement phase to the prediction procedure hence increasing reconfiguration time.

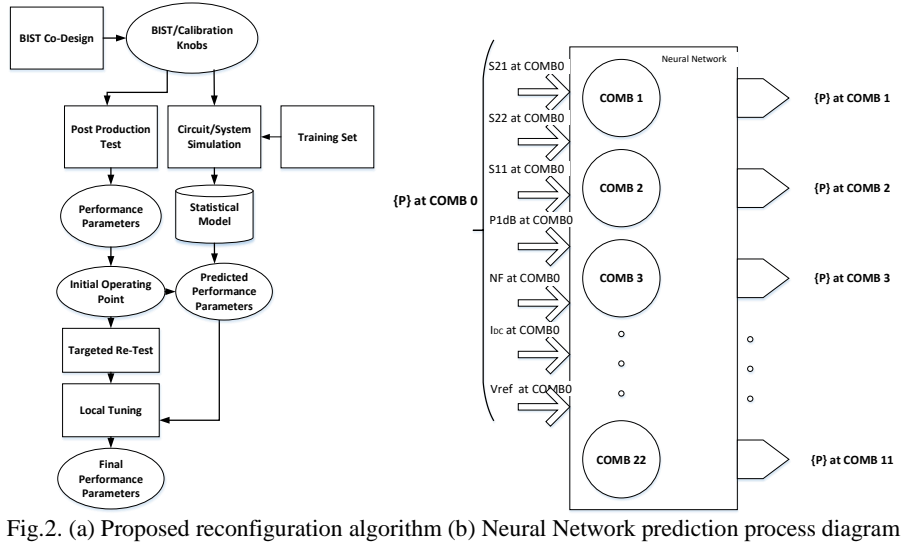

Fig.2. (a) Proposed reconfiguration algorithm (b) Neural Network prediction process diagram

## IV. LNA ARCHITECTURE

### A. LNA Design

Figure (3) shows the topology of the reconfigurable wide-band LNA. A current re-use technique is used to comply with the low voltage design. It provides high gain while driving high impedance of the second stage [19]. A DC feedback loop is used to define the operating points and keep $ML_{1A}$ and $ML_2$ in saturation [20]. $L_i$, $L_{s1}$ and $L_{s2}$ are tuned to obtain the optimum noise figure and input matching. The two-stage topology helps with the independent tuning of the performance parameters.

First stage primarily controls the noise figure while the second stage is mainly responsible for the linearity of the LNA. The gain control is conducted in three modes; High gain, medium gain and low gain, each obtainable with different combinations of noise figure and linearity. Hence, gain modes are available independent of the noise figure and linearity configurations.

The LNA is implemented in a 130 nm CMOS technology. The sizing of the transistors is listed in Table I. The varactors are added to compensate for bond wire inductances.

### B. Programmability Knob Selection

The tuning knobs include switches $SW_1$ to $SW_5$, $VDD1$ and $VDD2$ as it is shown in Fig. (3). The tuning hooks are selected such that they cover the desired reconfiguration range. $VDD1$ and $VDD2$ are supply voltages which are provided by a low-drop-out regulator since these knobs are controlled digitally too.

Five switches are embedded into the design to control the performance parameters. $SW_1$ connects gate of transistor $ML_{1B}$ to $ML_{1A}$ resulting in increase in gain and improving noise figure but aggregates input matching to some extent. $SW_2$ and $SW_3$ add a parallel resistance which consecutively changes the reference voltage of the common-mode feedback amplifier which affects the DC bias voltages of $ML_{1A}$, $ML_2$ and $ML_{3A}$. $SW_2$ reduces the reference voltage hence reduces gain, linearity and power consumption.

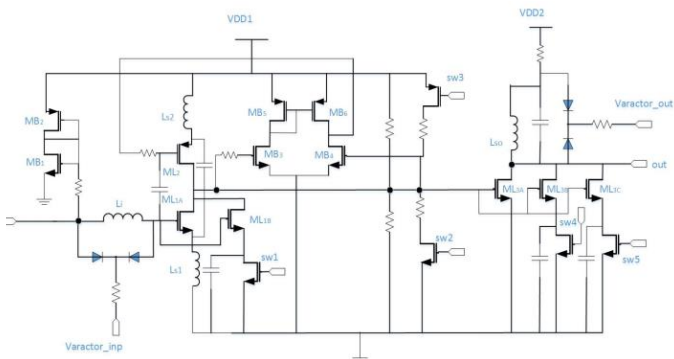

Fig.3. LNA circuit with embedded tuning knobs

*Table I. Device sizing*

| Device | $W/l\,(\mu m)$ | Device | $W/l\,(\mu m)$ |
|---|---|---|---|
| $ML_{1A}$ | 80/0.12 | $SW_2/SW_3$ | 2/0.12 |
| $ML_{1B}$ | 70/0.12 | $SW_1/SW_4/SW_5$ | 10/0.12 |
| $ML_2$ | 100/0.12 | $MB_1$ | 1/0.12 |
| $ML_{3A}$ | 30/0.12 | $MB_2$ | 2.5/0.12 |
| $ML_{3B}$ | 30/0.12 | $MB_3/MB_4$ | 3/0.12 |
| $ML_{3C}$ | 70/0.12 | $MB_5/MB_6$ | 8/0.12 |

Whereas, $SW_2$ increases the reference voltage hence increases linearity and power consumption but aggregates noise figure. $SW_4$ and $SW_4$ add transistors $ML_{3B}$ and $ML_{3C}$ in parallel with $ML_{3A}$ to resulting in increased gain and power consumption. Twelve programmable combinations are chosen to optimize performance with respect to requirements. For instance, if higher linearity is required, $SW_3$ is turned on however in order to reduce the noise figure $SW_1$ is switched on too. The desired performance can be achieved by setting the correct digital codes to the reconfiguration network.

## V. SIMULATION RESULTS

### A. Circuit Characterization

The conventional non-programmable LNA is characterized by running a 200-sample Monte-Carlo run. Fig. (4) depicts the distribution of each performance parameter.

The proposed reconfiguration scheme is used to sensitize the LNA circuit to the process variation. Hence, a Monte-Carlo simulation is performed to characterize the performance parameters range for each switching combination.

The performance corners are achieved by running Monte-Carlo simulation for 200 samples for all the switching combinations over process variation. Fig. (5) shows the histogram for each performance parameter of the programmable LNA. It reveals the sensitization effect on widening the circuit performance parameters span over process variation. Table II shows the tuning range for targeted performance parameters obtained from histogram plots. The reconfiguration feature provides a wide tuning range for gain, obtainable at a broad linearity span, makes it suitable for adaptation to localized application-specific requirements. Yet, noise figure variations is kept small providing the low noise figure requirement for the LNA.

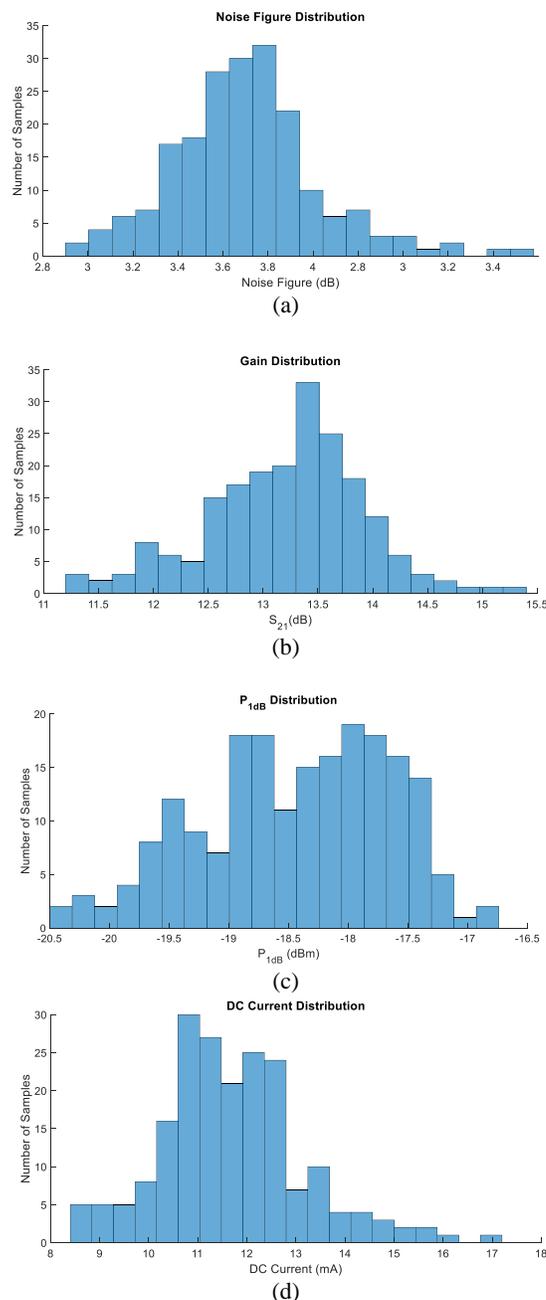

Fig.4. Distribution of a) Noise figure b) Gain c) $P_{1dB}$ d) DC current over process variation for conventional LNA

### B. Optimization in Matlab

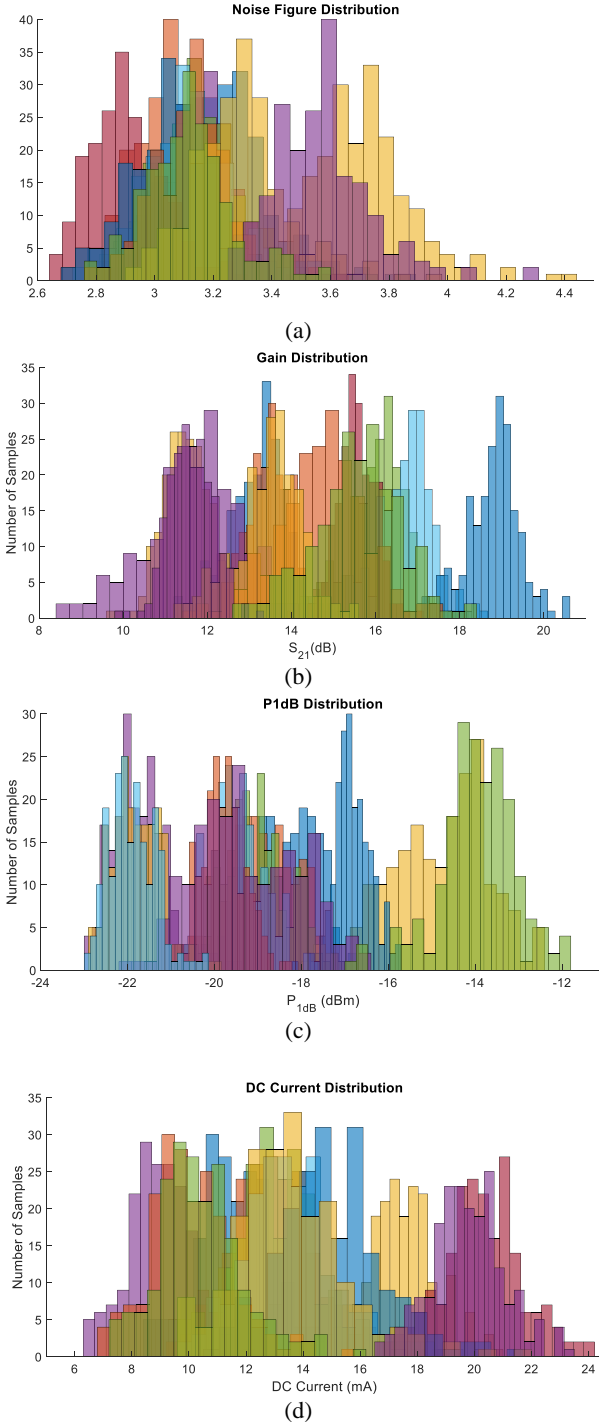

Fig.5. Distribution of a) Noise figure b) Gain c) $P_{1dB}$ d) DC current over process variation for programmable LNA

Twelve switching combination are used for the adaptation purpose.

As an example, different scenarios are investigated to show the adaptation of our design to the specific requirements. Fig. (6) shows different four scenarios where each has a particular performance parameter needs. Four different switching combination provides a close match for each situation. Table III lists the possible switch combinations which are the fits for each case.

As discussed earlier, the neural network learning algorithm is used in our proposed method to predict the performance parameters of the device. The prediction RMS error is calculated and plotted in Fig. (7) for gain, noise figure and P1dB. Maximum prediction RMS error for gain is 0.5 dB and for noise figure is 0.12dB. To illustrate the optimization and adaptation procedure, we review an example. A target performance is assumed: gain 15dB-17dB, $P_{1dB}$>-20dBm, NF<3.7dB. Three switching combinations satisfy the target requirements simultaneously as illustrated in Fig. (8). Therefore, the best switch combination is unknown. Instead of testing all three switch combinations, to save time, we apply the prediction algorithm. In this example, performance at combination 4 is characterized for a part and is fed to the algorithm as the known input to predict performance parameters at combination 5 and combination 6. Therefore, the test time is reduced to one-third of the conventional iterative approach where all potential switch combinations are being tested. However, there is a tradeoff between the adaptation time and the adaptation error. Alternatively, we may double the test time and characterize the DUT at two switching combinations and predict the third combination only. In this case, we assumed that the DUT performance at combination 4 and combination 6 are tested and are known; These are applied as the inputs to the prediction algorithm to predict the parameters at combination 5. After running our optimization algorithm, a switch combination which fits the desired performance the best with the optimum power consumption, is selected. The results for both cases are tabulated in Table IV. In this table measurement error is neglected. In case 1, the predicted gain for combination 5 is above the desired target range, hence it is removed from the choices. Between combination 4 and 6, combination 4 is more power efficient and can be the final choice.

*Table II. Tuning range of performance parameters*

| Performance Parameters | NF | P1dB | Gain | Power Diss. |
|---|---|---|---|---|
| Tuning Range | ≈1.5 dB | ≈10 dB | ≈12 dB | ≈18 mW |

*Table III. Four switch combinations for four different scenarios*

| Switch Combo | Gain (dB) | $S_{11}$ (dB) | $S_{22}$ (dB) | NF (dB) | $P_{1dB}$ (dBm) | $I_{DC}$ (mA) |
|---|---|---|---|---|---|---|
| Sw3 | 11.5 | -10.0 | -27.8 | 3.9 | -14.5 | 17.4 |
| Sw4-Sw5 | 19 | -11.4 | -25.6 | 3.2 | -20.0 | 15 |
| Sw1-Sw3-Sw4 | 16.5 | -9.9 | -30.7 | 3.0 | -16.5 | 20.3 |
| 1.05*VDD1-Sw2 | 13.5 | -14.4 | -26.5 | 3.5 | -19.2 | 13.5 |

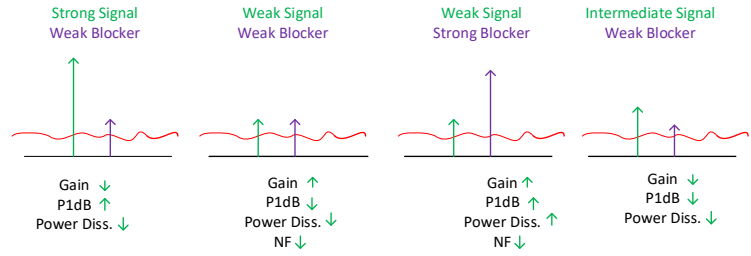

Fig. 6. Four scenarios for adaptation investigation

In case 2 on the other hand, predicted gain of combination 5 lies within the desired range. In this case, combination 5 can be the final choice due to lower power consumption.

Depending on the acceptable deviation from target parameters, an acceptable margin in target performance can be defined to account for the prediction error. Prediction error could be reduced by introducing more inputs to the prediction algorithm -shown in Fig.2- at the cost of more processing time.

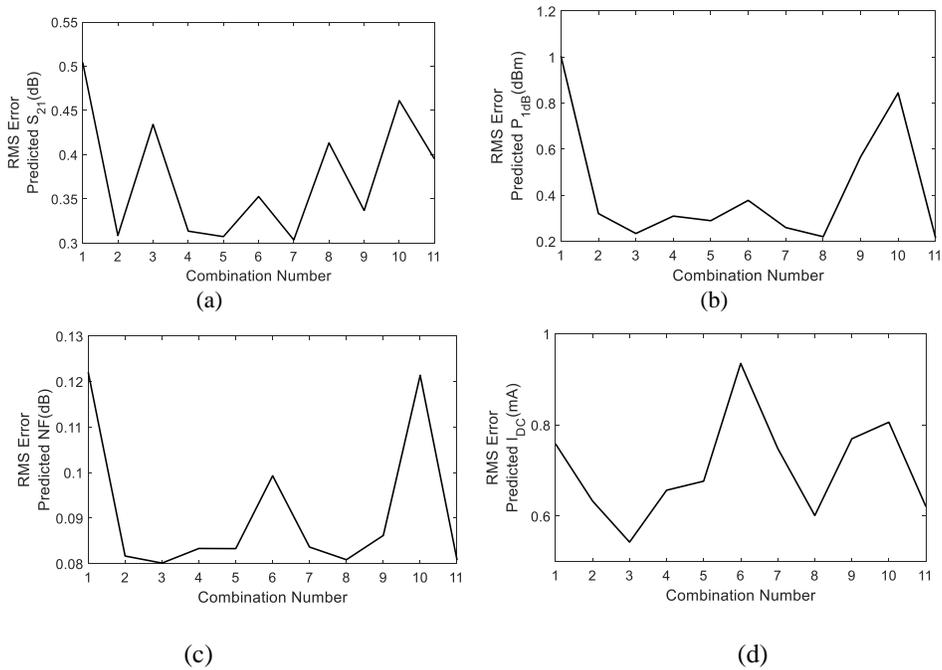

Fig.7 (a) Predicted gain RMS error (b) Predicted $P_{1dB}$ RMS error (c) Predicted noise figure RMS error (d) Predicted DC current RMS error

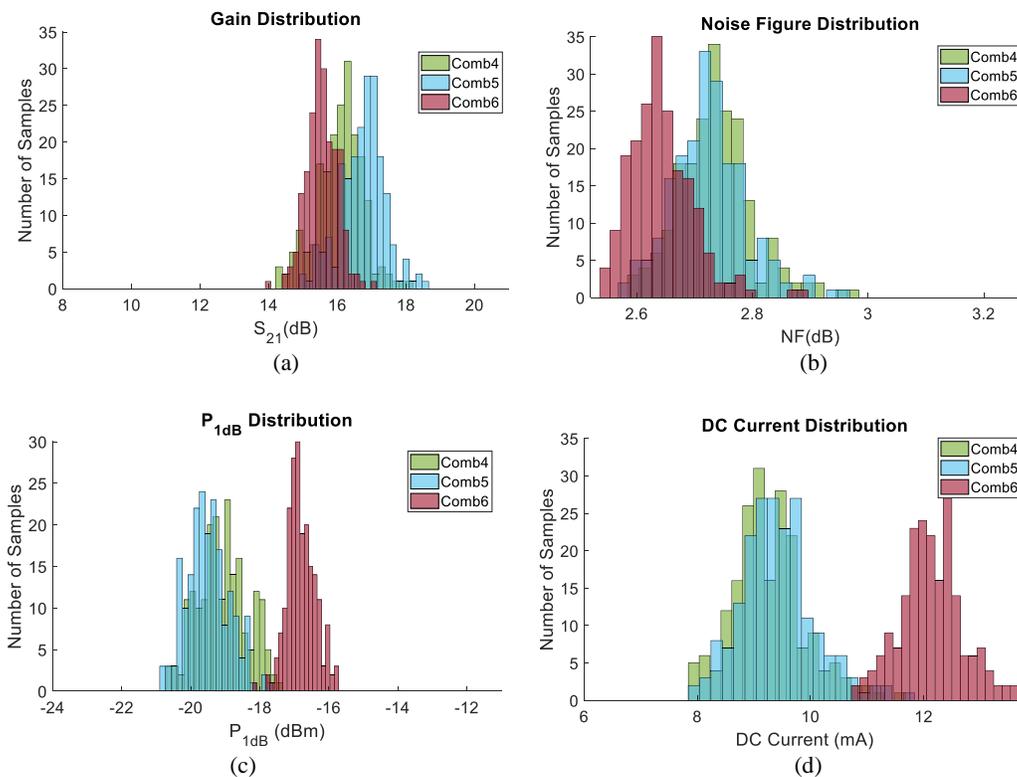

Fig.8. Performance parameters of three switching combinations over process variation (a) Gain (b) Noise figure (c) $P_{1dB}$ (d) DC current

## VI. CHIP IMPLEMENTATION

### A. Layout

The proposed adaptable LNA is designed in 130nm technology. Overall, the LNA occupies 0.16mm² area. The complete reconfiguration network, including the additional tuning modes occupies less than 0.0002mm² area. Thus, the entire reconfiguration network imposes no more than 0.1% area overhead. Fig. (9) depicts the layout and the microphotograph of the fabricated adaptable LNA including the pad ring and ESD cells. The distances between the on-chip inductors are optimized in a full-wave simulator to obtain the minimum unwanted coupling between the inductors.

*Table IV. Optimization result for specified target parameters*

| Target: $S_{21}$: 15dB-17dB $P_{1dB}$>-20dBm NF<3.7dB | | Combination 4 | | | | Combination 5 | | | | Combination 6 | | | |
|---|---|---|---|---|---|---|---|---|---|---|---|---|---|
| | | $S_{21}$(dB) | NF (dB) | $P_{1dB}$ (dBm) | $I_{DC}$ (mA) | $S_{21}$(dB) | NF (dB) | $P_{1dB}$ (dBm) | $I_{DC}$ (mA) | $S_{21}$(dB) | NF (dB) | $P_{1dB}$ (dBm) | $I_{DC}$ (mA) |
| Case1 | Measured | 16.29 | 3.65 | -18.45 | 15.0 | - | - | - | - | - | - | - | - |
| | Predicted | - | - | - | - | 17.13 | 3.54 | -18.87 | 15.0 | 15.73 | 3.27 | -16.7 | 20.8 |
| Case2 | Measured | 16.29 | 3.65 | -18.45 | 15.0 | - | - | - | - | 15.05 | 3.38 | -15.8 | 21.3 |
| | Predicted | - | - | - | - | 16.45 | 3.53 | -19.0 | 14.1 | - | - | - | - |
| Actual | | 16.29 | 3.65 | -18.45 | 15.0 | 16.85 | 3.62 | -18.95 | 14.0 | 15.05 | 3.38 | -15.8 | 21.3 |

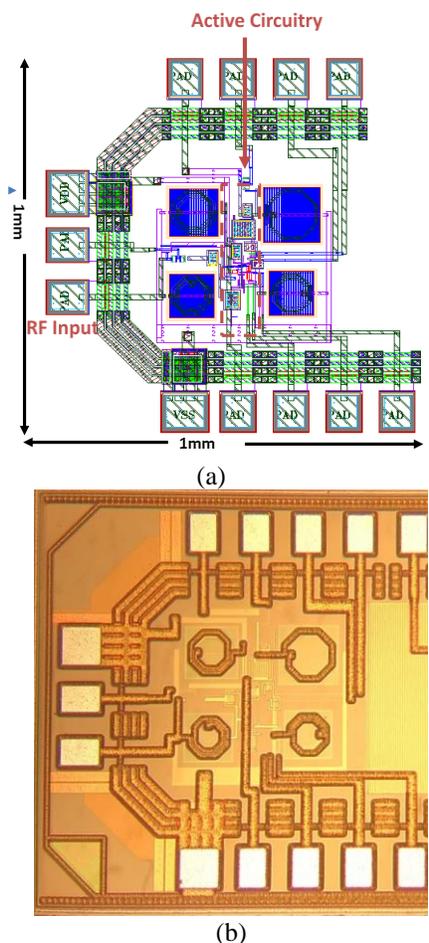

Fig. 9. (a) Proposed LNA chip layout in 0.13um process (b) The fabricated chip microphotograph

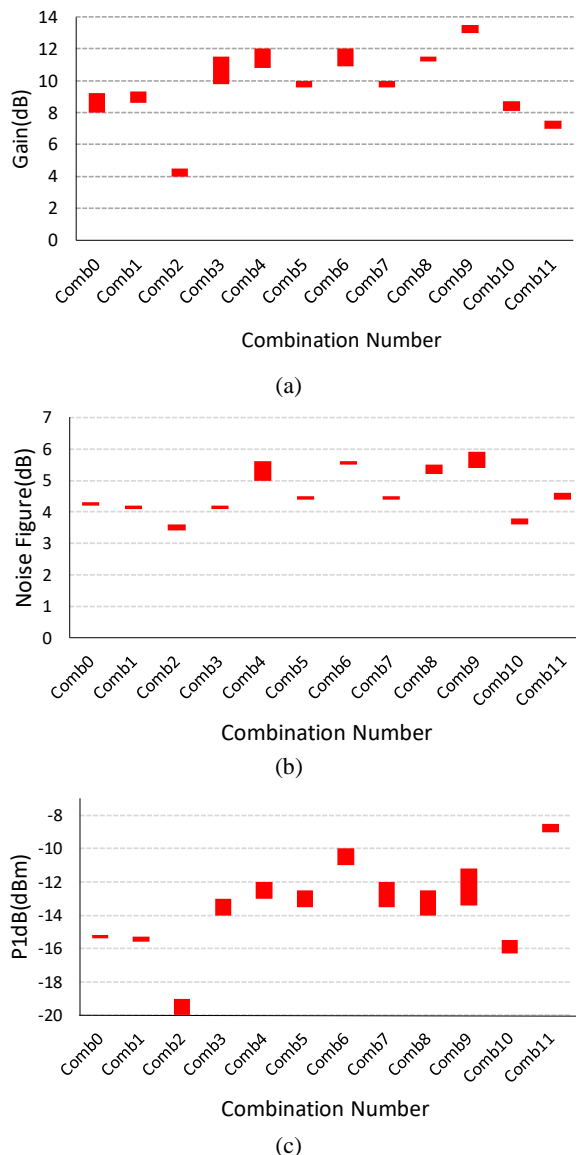

Fig.10. Performance parameters variation over four identical chips (a) gain (b) noise figure (c) $P_{1dB}$

## B. Chip Measurement Results

The fabricated chip is measured in the lab. A network analyzer is used to measure the S-parameters. A spectrum analyzer is used to measure P1dB of the device. To measure the noise figure, the Y-factor method is applied using a noise source and the spectrum analyzer. Four PCB boards with the LNA chip and same components are measured to account for the process variation. Fig.11 shows the variations in performance parameters for each switching combination. The variations among the boards are trained to generate a hundred Monte-Carlo samples. Using our learning algorithm in Fig.2, the RMS prediction errors for different switch combinations are obtained as shown in Fig.10. It is observed that the maximum gain prediction error is 0.3 dB, the maximum noise figure prediction error is 0.16dB and maximum $P_{1dB}$ prediction error is 0.8dB.

## VII. CONCLUSION

In this paper, we propose an automated adaptable sensor node for IOT applications. The machine learning technique is used for automatic adaptation. We demonstrated the concept by implementing it on a CMOS LNA with built-in tuning knobs. The performance range over process variation is inquired. Using the statistical model formed by learning algorithm over Monte-Carlo samples, the performance parameters are predicted. A case study of the in-field adaptation shows the effect of prediction error on the switching combination selection. By characterizing more combinations in the field and hence sacrificing the test time, a closer-to-target combination can be selected. The prediction algorithm applied to chip measurement results…

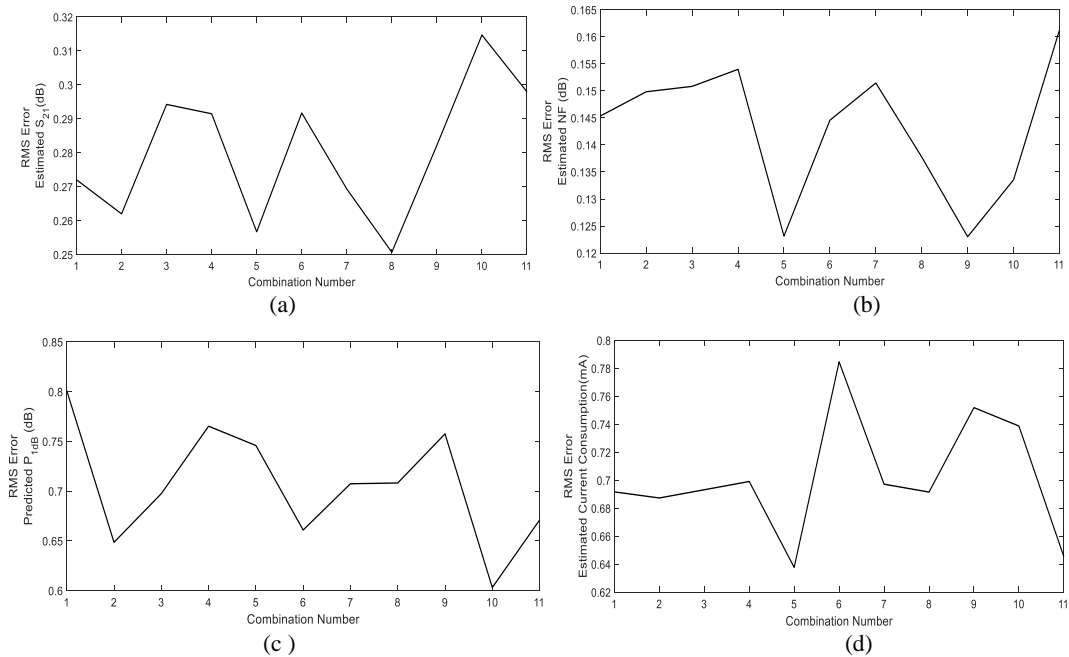

Fig.11. (a) Predicted gain RMS error (b) Predicted noise figure RMS error (c) Predicted $P_{1dB}$ RMS Error (d) Predicted current RMS error